\documentclass[11pt]{article}
\usepackage{moriond,epsfig}

\def\be{\begin{equation}}
\def\ee{\end{equation}}
\def\bea{\begin{eqnarray}}
\def\eea{\end{eqnarray}}

\begin{document}
\vspace*{4cm}

\title{DOUBLE CHOOZ: SEARCHING FOR $\theta_{13}$ WITH REACTOR NEUTRINOS}

\author{ P. NOVELLA }

\address{CIEMAT, Av. Complutense 22,\\
Madrid 28040, Spain}

\maketitle\abstracts{The discovery of neutrino oscillations is a direct indication of physics beyond the Standard Model. The so-called atmospheric and solar sectors have been explored by several experiments, meanwhile the mixing angle $\theta_{13}$ connecting both sectors remains unknown. In contrast to accelerator experiments, reactor neutrinos arise as a clean probe to search for this angle. The Double Chooz experiment is meant to search for $\theta_{13}$ taking advantage of the neutrinos generated at the nuclear power plant of Chooz. Double Chooz relies on neutrino flux measurements at two different locations, the so-called far an near detectors, although the first phase runs only with the far detector. The relative comparison of the fluxes at both sites will reduce the systematic uncertainties down to 0.6\%. The commissioning of the far detector took place between January 2011 and March 2011, when physics data taking started. First results are expected by the summer 2011. These results will improve the current limit to $\theta_{13}$ in case the oscillation signal is not observed. The final sensitivity to $\sin^{2}(2\theta_{13})$ is exepcted to be 0.03 at 90\% C.L. after 5 years of data taking.}

\section{One step beyond in neutrino oscillation physics}\label{sec:next}


Neutrino oscillation data can be described within a three neutrino mixing scheme, in which the flavor states $\nu_\alpha$ ($e,\nu,\mu$) are connected to the mass states $\nu_i$ ($i$=1,2,3) through the PMNS mixing matrix  $U_{PMNS}$ \cite{nurev}. This matrix can be expressed as the product of three matrices where the mixing parameters remain decoupled: $U_{PMNS} = U_{atm}\cdot U_{inter}\cdot U_{sol}$. The terms $U_{atm}$ and $U_{sol}$ describe the mixing in the so-called atmospheric and  solar sectors, which are driven by the mixing angles $\theta_{23}$ and $\theta_{12}$, respectively. The $U_{inter}$ matrix stands for the interference sector which connects the previous two, according to the mixing angle $\theta_{13}$ and the phase $\delta$ responsible for the $CP$ violation in the leptonic sector. Finally, the oscillation probability between two neutrino species becomes a function of the above oscillation parameters and the two independent mass squared differences $\Delta m^2_{ij}=m^2_i - m^2_j$.


The KamLAND experiment \cite{kam08} has explored the oscillation in the solar sector and provided allowed and best fit values for $\theta_{12}$ and $\Delta m^2_{21}$, showing consistency with solar experiments data. In the same way, the MINOS experiment \cite{minos} has published results for atmospheric sector ($\theta_{23}$ and $|\Delta m^2_{31}|$), being consistent with atmospheric neutrino data. However, the subdominant oscillation corresponding to the interference sector has not been observed yet. Results from CHOOZ experiment \cite{chooz} show at 90\% C.L. that $\sin^2(2\theta_{13})< 0.15$ for $|\Delta m^2_{31}|=2.5\times 10^{-3}$eV$^2$. Provided that $\delta$ appears in $U_{inter}$ only in combination with $\sin^2(2\theta_{13})$, the CP-violating phase also remains unknown. As a direct consequence, the search for the third mixing angle stands as one of the major open issues in neutrino oscillation physics.

\section{Reactor neutrinos in the quest for $\theta_{13}$}\label{sec:reactor}


Nuclear reactors produce nearly pure $\bar\nu_e$ fluxes coming from $\beta$ decay of fission fragments. A typical core delivers about $2\times 10^{20}$ $\bar\nu_e$ per second and GW$_{th}$ of thermal power. Such high isotropic fluxes compensate for the small neutrino cross-section and allow for an arbitrary location of neutrino detectors, scaling the flux with $1/L^{2}$ where $L$ is the distance between the core and the detector. Any oscillation effect in the $\bar\nu_e$ survival is governed by the following equation:

\begin{equation}
\label{eq:p}
P(\bar{\nu}_e \rightarrow \bar{\nu}_e) \cong 1
- \sin^2 2 \theta_{13} \sin^2(\frac{\Delta m^2_{31} L}{4E_{\nu}})  
- \cos^4 \theta_{13} \sin^2 2 \theta_{12} 
\sin^2(\frac{\Delta m^2_{21} L}{4E_{\nu}}) 
\end{equation} 


\noindent where $E_{\nu}$ is the neutrino energy. The second and third terms of Eq.~\ref{eq:p} describe the oscillation driven by $\theta_{13}$ and $\theta_{12}$ (solar regime), respectively. The value of $\theta_{13}$ can be derived directly by measuring $P(\bar{\nu}_e \rightarrow \bar{\nu}_e)$. Notice that in contrast to accelerator neutrino experiments, this measurement does not suffer from the $\delta-\theta_{13}$ degeneracy.        


\subsection{Detecting reactor neutrinos}

\begin{figure}[htbp]
\begin{center}
\includegraphics[width=60mm]{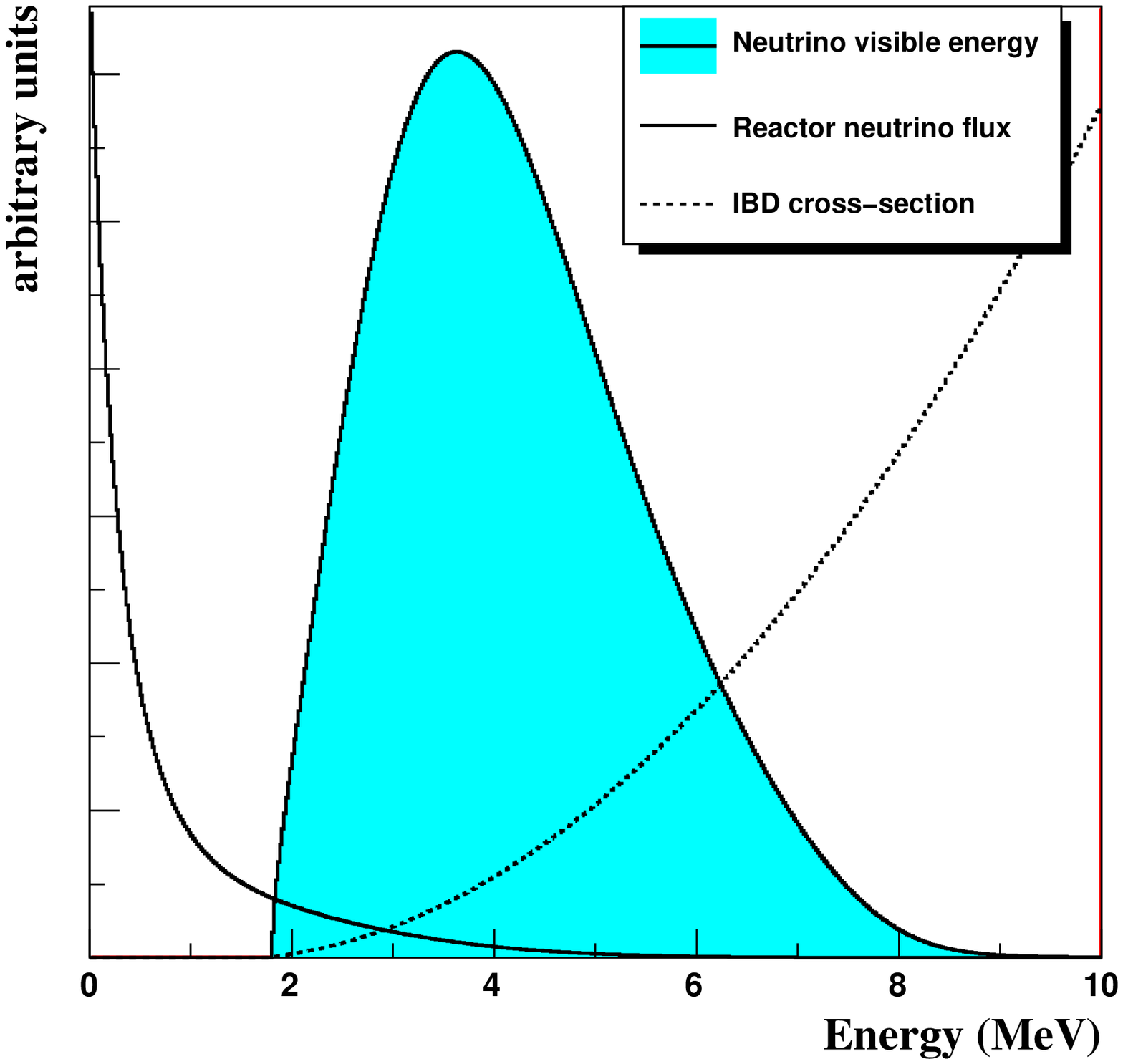}
\includegraphics[width=60mm]{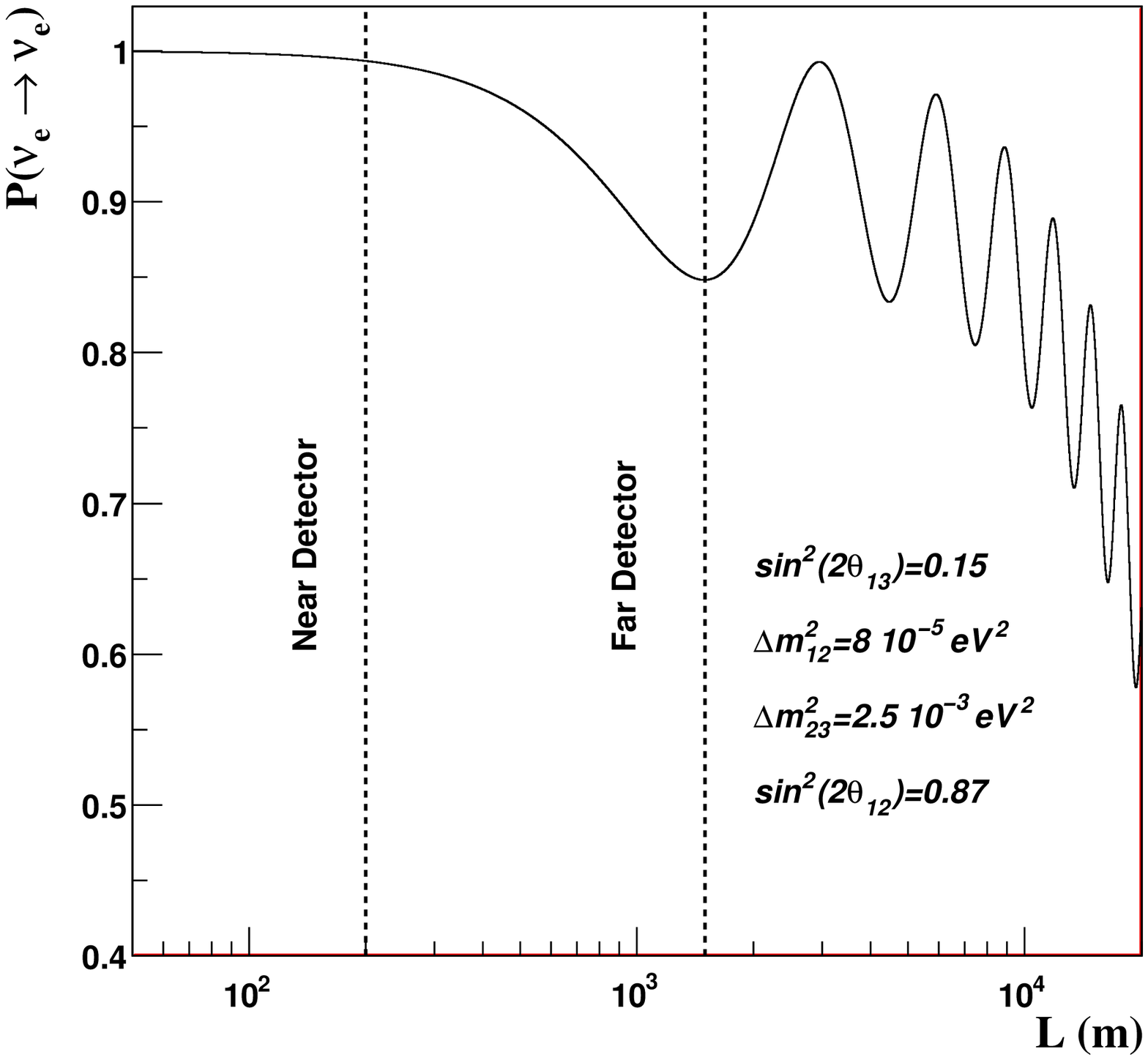}
\end{center}
\caption{\label{fig:nuexp} Left: $\bar\nu_e$ visible spectrum as a result of the flux shape and IBD cross-section. Right: $\bar\nu_e$ survival probability for $E_{\nu}=3$ MeV, as a function of the distance $L$.}
\end{figure}

The most common way of detecting reactor neutrinos is via the inverse beta decay (IBD) $\bar\nu_e +p \rightarrow n + e^+$. When this reaction takes place in liquid scintillator doped with Gadolinium, it produces two signals separated by about $\sim30$ $\mu$s: the first one due to the $e^+$ and its annihilation (prompt signal), and the second one due to the $n$ capture in a Gd nucleus (delayed signal). This characteristic signature yields a very efficient background rejection. The $e^+$ energy spectrum peaks at $\sim 3$MeV and can be related to $E_{\nu}$. The mean energy of the $\bar\nu_e$ spectrum in a detector filled by such a scintillator is around $4$ MeV, as shown in left panel of Fig.~\ref{fig:nuexp}. According to Eq.~\ref{eq:p}, for this energy the oscillation effect due to $\theta_{13}$ starts to show up at $L\sim0.5$ km, where the effect of $\theta_{12}$ is still negligible. Therefore, neutrino reactor experiments with short baselines become a clean laboratory to search for $\theta_{13}$.     



In spite of its characteristic signature, the IBD signal can be mimicked by the so-called accidental and correlated backgrounds. The accidental background is defined as the coincidence of a positron-like signal coming from natural radioactivity, and the capture in the detector of a neutron created by cosmic muon spallation in the surrounding rock. The correlated background consists of events which may mimic both the prompt and the delayed signals of the IBD. Fast neutrons and cosmogenic isotopes, both generated in muon interactions, are the main sources of this background. Fast neutrons are produced by muons in the surrounding rock and enter the detector leading to proton recoils, thus faking a prompt signal, before being captured by a Gd nucleus. Muons also produce inside the detector long-lived $\beta$-n decay isotopes, like $^9$Li and $^8$He. As the half-life of such cosmogenic isotopes is $\sim$100 ms, their decay cannot be related to the muon interaction.

\section{Getting the most from reactor experiments}\label{sec:optim}

The sensitivity to the $\theta_{13}$-driven oscillation is optimized by detecting a deficit in the expected neutrino events around 1 km away from the nuclear power plant, as shown in right panel of Fig.~\ref{fig:nuexp}. However, some of the largest systematics in reactor experiments arise from the uncertainties in the original $\bar\nu_e$ fluxes. In order to reduce them, a relative comparison between two or more identical detectors located at different distances from the reactors becomes critical. In particular, a detector placed a few hundred meters away can measure the fluxes before any oscillation takes place, as demonstrated in right panel of Fig.~\ref{fig:nuexp}. The comparison between the so-called far and near detectors leads to a breakthrough in the sensitivity to $\theta_{13}$, as all the fully correlated systematics cancel out. Further steps in the sensitivity optimization relay on reducing the relative normalization and the relative energy scale uncertainties of the detectors, as well as on minimizing the backgrounds.   

\section{The Double Chooz approach}\label{sec:dchooz}

The Double Chooz experiment \cite{dcprop}, located at the nuclear power plant of Chooz (France),  aims at improving the CHOOZ experience by means of a long-term stability multi-detector setup. The comparison between un-oscillated reactor neutrino flux at a near site  and the oscillated flux at a far site allows for the cancellation of the reactor-related correlated errors. The detector-related systematics are kept under control by constructing two identical detectors providing accurate energy reconstruction and high signal-to-noise ratios. The Double Chooz collaboration involves institutes from Brazil, France, Germany, Japan, Russia, Spain, UK and USA. 

The Chooz nuclear plant consists of two cores yielding a total thermal power of 8.54 GW$_{th}$. The Double Chooz far detector is placed 1050 m away from the cores, in the same underground laboratory used by the CHOOZ experiment. The laboratory is located close to the maximal oscillation distance and provides enough shielding (300 m.w.e.) against cosmic rays. A second identical detector (near detector) will be installed 400 m away from the reactor cores, in a new laboratory (115 m.w.e) whose construction started in April 2011. 

\subsection{The Double Chooz detectors}\label{sec:detectors}

\begin{figure}[htbp]
\begin{center}
\includegraphics[width=100mm,height=70mm]{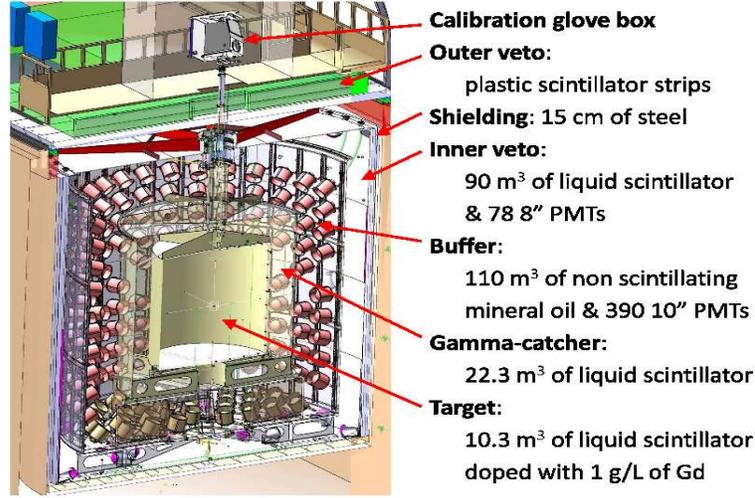}
\end{center}
\caption{\label{detector}The Double Chooz detector design.}
\end{figure}

The Double Chooz detectors design is optimized to reduce backgrounds. The detectors, shown in Fig.~\ref{detector}, consist of a set of concentric cylinders and an outer plastic scintillator muon veto ({\em outer veto}) on the top. The innermost volume ({\em target}) contains about 10 tons of Gd-loaded (0.1\%) liquid scintillator inside a transparent acrylic vessel, where the neutrinos interact via the IBD process. This volume is surrounded by another acrylic vessel filled with unloaded scintillator ({\em gamma-catcher}). This second volume is meant to fully contain the energy deposition of gamma rays from the neutron capture on Gd and the positron annihilation in the target region. The gamma-catcher is in turn contained within a third volume ({\em buffer tank}) made of stainless steel and filled with mineral oil. As the wall and the lids of the buffer are covered with an array of 390 10'' photomultiplier tubes (PMTs), meant to detect the scintillation light (13\% photocathode coverage), the oil shields the target and the gamma-catcher against the radioactivity of the PMT components. The target, gamma-catcher and buffer tank define the {\em inner detector}. Finally, the outer volume containing the inner detector is a stainless steel vessel covered with 78 8'' PMTs and filled with scintillator. This volume plays the role of the {\em inner muon veto}. To protect the Double Chooz detector from the radioactivity of the surrounding rock, a 15 cm layer of iron is used.    

The detector performance is analyzed by means of a redundant set of calibration systems. Apart from the natural calibration sources (neutron captures in H, Gd and C), radioactive sources can be introduced in the different volumes of the detector, via a glove box. The goal is to achieve a relative error on the neutrino detection efficiency of 0.5\% with both detectors, and an energy scale uncertainty of 0.5\%. In addition, a set of LEDs embedded in the PMTs structure is used to measure the PMTs gains and timing, as well as to monitor the stability of the detector.   

\subsection{Improving the CHOOZ experience}

The Double Chooz experiment aims at improving the CHOOZ result by means of an increase of the exposure and a reduction of the systematics. In order to reduce the statistical error down to 0.5\% (was 2.8\% in CHOOZ), a long-term stability scintillator has been developed, which will allow for a total data taking time of 5 years. Besides, a larger target volume (10.3 m$m^3$) is used. The relative comparison of the fluxes between the far and the near detector will allow for the reduction of the systematics error down to 0.6\% (was 2.7\% in CHOOZ), as shown in Tab. \ref{tab:sys}. Backgrounds are also expected to be reduced with respect to CHOOZ due to the selection of radiopure materials used in the detector, the two independent muon vetoes, and the buffer volume which isolates the PMTs from the active part of the detector.



\begin{table}[htbp]
\caption{\label{tab:sys} Main systematic uncertainties in CHOOZ and Double Chooz reactor experiments.} 
\begin{center}
\begin{tabular}{|c|c|c|}
\hline
  & \textbf{CHOOZ} & \textbf{Double Chooz}\\
\hline
Reactor fuel cross section & 1.9\% & -- \\
Reactor power & 0.7\% & -- \\
Energy per fission & 0.6\% & -- \\
Number of protons & 0.8\% & 0.2\% \\
Detection efficiency & 1.5\% & 0.5\% \\
\hline
\textbf{TOTAL} & 2.7\% & 0.6\% \\
\hline
\end{tabular}
\end{center}
\end{table}



\section{The newborn detector}\label{sec:fardetector}

\begin{figure}[htbp]
\begin{center}
\includegraphics[width=80mm]{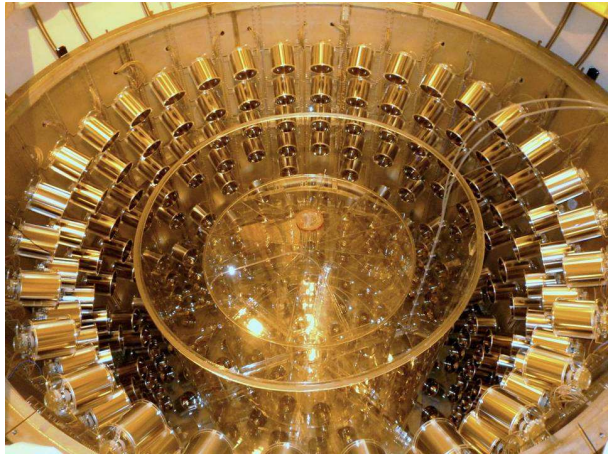}
\end{center}
\caption{\label{fig:vista} View of the acrylic vessels and the PMTs covering the the buffer tank walls.}
\end{figure}

The integration phase of the far detector of Double Chooz started in May 2008 with the integration of the external shield, followed by the assembly of the inner veto tank. The buffer vessel was completed in summer 2009, and the 330 10'' PMTs were successfully mounted by fall 2009. Finally, the acrylic gamma-catcher and target vessels were installed inside the buffer, as shown in Fig.~\ref{fig:vista}, and the detector was closed. First signals from the PMTs were observed in summer 2010 as the DAQ and electronics systems became ready. The filling of the detector started in October 2010 and was completed by the end of 2010. 

\begin{figure}[htbp]
\begin{center}
\includegraphics[width=70mm]{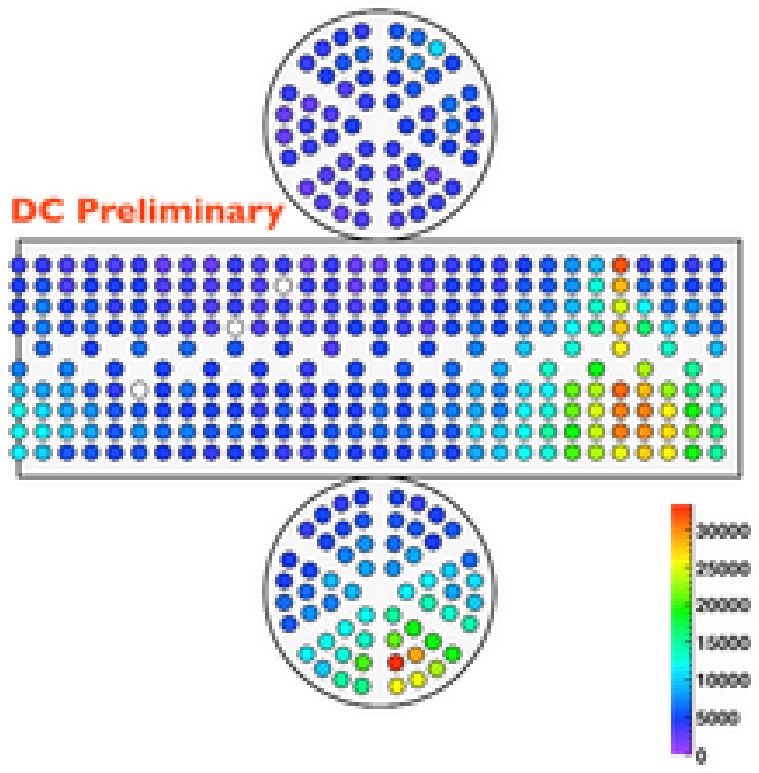}
\includegraphics[width=70mm]{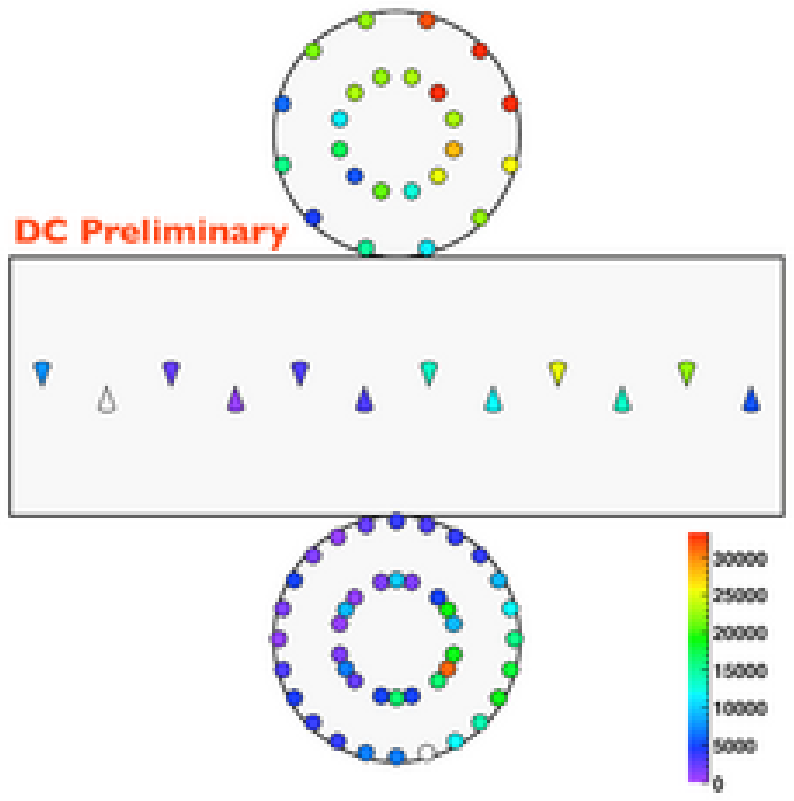}
\end{center}
\caption{\label{fig:display} Display of a crossing muon event in both inner detector (left) and inner veto (right). Colors show the charge (digital units) collected at each PMT.}
\end{figure}

The commissioning period took place between January 2011 and March 2011. First analysis of the detector response was carried out, assuring the good performance of both the inner detector and the inner veto. First events in the filled detector were observed in January 2011. As an example, the display of a muon crossing both detectors is shown in Fig.~\ref{fig:display}.

\section{Getting the most from Double Chooz}\label{sec:sens}

\begin{figure}[htbp]
\begin{center}
\includegraphics[width=100mm]{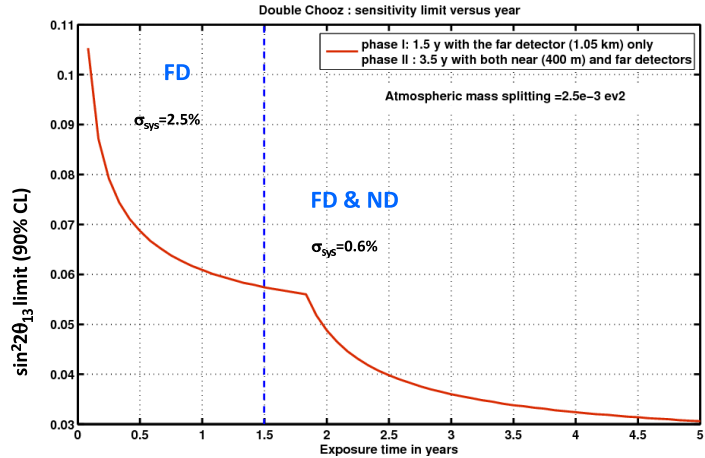}
\end{center}
\caption{\label{fig:sensi}Double Chooz expected sensitivity limit (90\% C.L.) to $\sin^2(2\theta_{13})$ as a function of time for $\Delta$m$^2_{31}$ = 2.5 $\times$ 10$^{-3}$ eV$^2$. Near detector is assumed to be ready 1.5 years after the start of the far detector operation.}
\end{figure}

The Double Chooz experiment is developed in two phases. Phase I started in March 2011 once the commissioning of the far detector was completed. Even operating only one detector, this phase will be able to improve the current $\theta_{13}$ limit in a few months of data taking. A sensitivity of $\sin^{2}(2\theta_{13}) \sim 0.6 $ is expected after 1.5 years of data taking. The oscillation analysis in phase I will be limited by the uncertainties in reactor fluxes, being the total systematics around 2.8\%. Ultimate systematics reduction, down to 0.6\%, will be achieved in Phase II (2012) when the second detector (near site) starts taking data. Dominant errors in this phase will be the relative detector normalization and energy scale and the event selection cuts. After 4 months of data taking, Phase II will improve the results of Phase I. After 3 years of data taking, a sensitivity to $\sin^{2}(2\theta_{13})$ of 0.03 (90\% C.L.) will be achieved. A 3$\sigma$ measurement will be feasible if $\sin^{2}(2\theta_{13}) > 0.05$. A summary of the sensitivity of both phases is shown in Fig.~\ref{fig:sensi}.

\vspace{0.5cm}

\end{document}